\documentstyle[11pt,epsfig]{article}

\newcommand{\ds}{\displaystyle }
\newcommand{\R}{{\sf R\hspace*{-0.94ex}%
\rule{0.15ex}{1.5ex}\hspace*{0.94ex}}}

\newcommand{\N}{{\sf N\hspace*{-0.96ex}%
\rule{0.15ex}{1.5ex}\hspace*{0.96ex}}}
\newcommand{\C}{{\sf C\hspace*{-0.96ex}%
\rule{0.15ex}{1.5ex}\hspace*{0.96ex}}}

\title{Sur une g\'en\'eralisation de l'op\'erateur fractionnaire}

\author{ {\sl Thomas M.  Michelitsch$^{1}$\footnote{Auteur correspondant, couriel~: michel@lmm.jussieu.fr}, G\'{e}rard A. Maugin$^{1}$}\\ \\ {\sl Shahram Derogar$^2$} \\ \\ {\sl Andrzej F. Nowakowski$^3$, Franck C. G. A. Nicolleau$^3$} \\ \\
$^1$ Universit\'{e} Pierre et Marie Curie, Paris 6\\
Institut Jean le Rond d'Alembert \\ CNRS UMR 7190 \\ France
 \\
$^2$ School of Mechanical, Aerospace and Civil Engineering\\
The University of Manchester
\\
Royaume Uni
\\
$^3$ Sheffield Fluid Mechanics Group, Department of Mechanical Engineering\\
University of Sheffield\\
Royaume Uni
\\
\\
}

\begin{document}

\maketitle

\noindent {\it paru dans des actes de congr\` es, CFM11,\\ Besan{\c c}on,
le 20\` eme CONGRES FRANCAIS DE MECANIQUE DU 28 AOUT AU 2 SEPTEMBRE 2011
Communication no. 75.}

\paragraph{R\'esum\'e}
Cette communication a pour but de proposer une g\'en\'eralisation de la notion de d\'eriv\'ee traditionnelle afin d'inclure les d\'eriv\'ees fractionnaires comme
celles de Riemann-Liouville, Gruenwald-Letnikov, Weyl, Riesz, Caputo, Marchaud
et d'autres variantes comme cas particuliers. Nous le d\'emontrons de mani\`ere explicite pour la d\'eriv\'ee de Marchaud.
Afin de d\'efinir cette notion il convient d'employer le calcul d'op\'erateurs lin\'eaires. La notion de d\'eriv\'ee g\'en\'eralis\'ee est susceptible non seulement de reproduire les d\'eriv\'ees et int\'egrales
traditionelles et fractionnaires mais aussi de d\'efinir de nouvelles esp\`eces
de ``d\'eriv\'ees" qui peuvent \^etre utiles pour traiter certains probl\`emes
de m\'ecanique en milieux sans echelles internes (fractals) \cite{michel-phys_rev_e}.

\paragraph{mots cl\'e:} D\'eriv\'ee g\'en\'eralis\'ee, d\'eriv\'ee fractionnaire, int\'egrale fractionnaire de Riemann-Liouville, diff\'erence fractionnaire, calcul d'op\'erateurs

\section{INTRODUCTION}

Le but du calcul fractionnaire qui a environ 300 ans est d'appr\'ehender le probl\`eme des ordres non-entiers des d\'eriv\'ees
traditionnelles. Comme il est bien connu, l'extension de la notion de d\'eriv\'ee aux ordres non-entiers ne se fait pas de mani\`ere unique.
Du coup, il en existe plusieurs variantes. Nous nous
r\'ef\'erons \`a l'ouvrage classique d'Oldham et Spanier \cite{oldham} et \`a d'autres auteurs dont nous ne citons que \cite{ross,miller,riesz,samko}.

Cette communication a pour but d'introduire une g\'en\'eralisation des notions de ``d\'eriv\'ee" et d'``int\'egrale" afin de mettre en place un outil math\'ematique permettant d'aborder quelques probl\'emes nouveaux en m\'ecanique des mat\'eriaux auto-similaires. Pour cela il convient
d'employer l'op\'erateur de translation $T$ d\'efini par
\begin{equation}
\label{shiftop}
T(h)f(x)=f(x+h) \, \,, \forall x,h \in \R
\end{equation}

Pour \'eviter trop de complications nous nous restreignons ici \`a des fonctions continues, soit plusieurs fois diff\'erentiables, soit non-diff\'erentiables. On observe que $T(h_1)T(h_2)=T(h_1+h_2),\, h_1,h_2 \in \R$ avec $T(0)=1$ et $T^a(h)=T(ah)\,, a\in \R$.
Si $f(x)$ est sufisamment lisse (lipschitzienne) on peut identifier
\begin{equation}
\label{shiftexp}
T(h)=e^{hD_x}
\end{equation}
 o\`u $D_x=\frac{d}{dx}$ d\'enote la d\'eriv\'ee traditionnelle du premier ordre. Dans cette communication nous construisons des op\'erateurs qui s'appliquent aux fonctions continues remplissant la condition de
 H\"{o}lder \cite{ross}:

\begin{equation}
\label{hoeldercondition}
|f(x+h)-f(x)| \leq C \, h^{\delta} \,\,,\hspace{2cm} 0 < \delta \leq 1
\end{equation}
pour $x$ et $x+h$ \'etant dans le domaine de $f(x)$. $C$ d\'esigne une constante et $\delta$ l'exposant de H\"{o}lder. Si $\delta\neq 1$ la fonction $f(x)$ n'est pas diff\'erentiable et par cons\'equent (\ref{shiftexp}) n'existe pas. Le cas $\delta=1$ correspond aux fonctions traditionnelles dites liptschitziennes.

\section{Motivation et d\'efinition de d\'eriv\'ees et int\'egrales g\'eneralis\'ees}
\label{defgender}

Cet paragraphe est consacr\'e \`a une g\'en\'eralisation de la notion de d\'eriv\'ee incluant la notion
de d\'eriv\'ee fractionnaire. Il en r\'esulte un op\'erateur qui s'applique aux fonctions {\it h\"{o}ld\'eriennes}. Pour motiver notre propos, consid\'erons
la d\'eriv\'ee d'ordre $n$ entier ($n \in \N_0$) d\'efinie par
\begin{equation}
\label{limprocdern}
D_x^nf(x)=\lim_{h\rightarrow 0}h^{-n}\left(T(h)-1\right)^nf(x)=
\lim_{h\rightarrow 0}h^{-n}\sum_{k=0}^n (-1)^k\left(\begin{array}{c}\!\!\! n \!\!\! \nonumber \\ \!\!\! k \!\!\! \end{array}\right)f(x-kh) \,,\hspace{2cm} n \in \N_0
\end{equation}
o\`u $\lim_{h\rightarrow 0}T(nh)\rightarrow 1$ est utilis\'e.
Dans (\ref{limprocdern}) on a utilis\'e les notations habituelles $\left(\begin{array}{c}\!\!\! n \!\!\! \nonumber \\ \!\!\! k \!\!\! \end{array}\right)=\frac{n!}{k!(n-k)!}$ avec la fonction  $\Gamma$ d\'efinie par\footnote{Il nous suffira ici de nous restreindre aux $\beta \in \R$.}
\begin{equation}
\label{betafaculty}
\beta!=\Gamma(\beta +1)=\int_0^{\infty}\tau^{\beta}e^{-\tau}{\rm d}\tau \,\,, \beta \in \R > -1
\end{equation}
g\'en\'eralisant $\beta !$ pour les non-entiers verifiant\footnote{$Re(\beta) > -1$ si $\beta \in \C$.} $\beta > -1$ et qui donne pour $\beta=n=1,2,..\in \N$
$n!=\Pi_{k=1}^n k$. La restriction $\beta >-1$ assure la convergence de l'int\'egrale (\ref{betafaculty}).
Au vu de (\ref{limprocdern}) la d\'eriv\'ee traditionnelle d'ordre $n$ peut \^etre r\'e-\'ecrite comme

\begin{equation}
\label{limprocdern1}
D^nf(x)=\lim_{h\rightarrow 0} \frac{g(T(h))}{g(1+h)}f(x)
\end{equation}
avec une fonction

\begin{equation}
\label{gfuexample}
g(\lambda)=(\lambda-1)^n=\sum_{k=0}^{\infty}a(k)\lambda^k    ,\hspace{1.5cm}a(k)=(-1)^{n-k}\left(\begin{array}{c}\!\!\! n \!\!\! \nonumber \\ \!\!\! k \!\!\! \end{array}\right)
\end{equation}
o\`u nous ne nous int\'eressons qu'aux cas non-banals $n \geq 1 \in \N$. Nous d\'enommons par la
suite la fonction $g(\lambda)$ ``fonction constituante" ou en bref la ``constituante".
On observe dans l'exemple ci-dessus que

\begin{equation}
\label{exprop1}
g(\lambda=1)=\sum_{k=0}^{\infty}a(k)=0
\end{equation}

A partir de cette observation nous appelons le cas limite

\begin{equation}
\label{limproc2}
{\cal D}_{g}f(x)=\lim_{h\rightarrow 0} \frac{g(T(h))}{g(1+h)}f(x)
\end{equation}
``{\it d\'eriv\'ee} g\'eneralis\'ee", en bref ``d\'eriv\'ee", si sa constituante satisfait (\ref{exprop1}), c'est-\`a-dire si
$g(\lambda=1)=0$ et par cons\'equent la d\'eriv\'ee g\'en\'eralis\'ee d'une constante est z\'ero. L'exemple le plus banal de cette notion est bien s\^ur la d\'eriv\'ee
traditionnelle du premier ordre ayant la constituante $g(\lambda)=\lambda -1$.
Inversement nous appelons (\ref{limproc2}) ``{\it int\'egrale} g\'en\'eralis\'ee" ou simplement ``int\'egrale" si $g(\lambda=1)$ est singuli\`ere au point $\lambda=1$ \`a savoir
\begin{equation}
\label{exprop2}
|\lim_{\lambda \rightarrow 1} g(\lambda)=\sum_{k=0}^{\infty}a(k)| \rightarrow \infty
\end{equation}
Il s'ensuit que si une constituante $g(\lambda)$ r\'ealise une d\'eriv\'ee, alors $1/g(\lambda)$ r\'ealise
une int\'egrale et vice versa. Pour tous les autres cas o\`u $0<|g(1)|<\infty$, (\ref{limproc2}) peut \^etre d\'etermin\'ee sans probl\`eme
donnant le r\'esultat banal ${\cal D}_gf(x)=f(x)$.
Dans cette communication nous nous restreignons aux cas o\`u la constituante
a un comportement au voisinage de $\lambda=1$ comme une puissance~: $(\lambda-1)^c$, $c\in \R$. Dans ce cas on peut \'egalement remplacer (\ref{limproc2}) par ${\cal D}_gf(x)=\lim_{h\rightarrow 0} g(1+[T(h)-1]/h)$.

D'abord nous donnons une \'evaluation qui est valable pour les cas o\`u $\frac{d^s}{d\lambda^s}g(\lambda =1) < \infty , s\in \N_0$ et au moins d'ordre $s_0$ avec $\frac{d^s}{d\lambda^s}g(\lambda =1)\neq 0$.
Pour cela nous posons

\begin{equation}
\label{gderivt}
g(T(h))=g(\lambda+T(h)-1)|_{\lambda=1}=e^{(T(h)-1)\frac{d}{d\lambda}}g(\lambda)|_{\lambda=1}
\end{equation}
ce qui peut \^etre \'ecrit en \'evaluant l'op\'erateur exponentiel comme

\begin{equation}
\label{recritrel}
\begin{array}{lll}
\ds g(T(h))f(x)=\sum_{s=0}\frac{1}{s!}(T(h)-1)^s\frac{d^s}{d\lambda^s}g(\lambda)|_{\lambda=1} f(x)&& \nonumber \\ && \nonumber \\
\ds = g(1)f(x) +(T(h)-1)f(x)\frac{d}{d \lambda}g(\lambda=1)+\frac{1}{2!}(T(h)-1)^2f(x)\frac{d^2}{d \lambda^2}g(\lambda=1)+..&& \nonumber \\ &&
\end{array}
\end{equation}
Par la suite tous les d\'ependances de $h$ se comprennent dans le cas limite $h\rightarrow 0$ ($h>0$)\footnote{Sans que "$h\rightarrow 0$" soit toujours explicitement \'ecrit.}.
Au cas o\`u $f(x)$ serait infiniment diff\'erentiable on peut remplacer $(T(h)-1)^sf(x)=(e^{hD_x}-1)^sf(x)=h^sD_x^sf(x)$. Le cas limite $h\rightarrow 0$ de la s\'erie  (\ref{recritrel})
est donc d\'etermin\'e par l'ordre $s_0$ de d\'erivation de $g$ le plus bas donnant une d\'eriv\'ee non-nulle, soit~:
\begin{equation}
\label{nulcond}
\frac{d^s}{d \lambda^s}g(\lambda=1) = 0 ,\, \forall \, 0\leq s < s_0
\end{equation}
c'est \`a dire que $0 < |\frac{d^{s_0}}{d \lambda^{s_0}}g(\lambda=1)| < \infty$.
Il s'ensuit que l'ordre dominant dans (\ref{recritrel}) pour $h\rightarrow 0$  est donc l'ordre
$s_0$

\begin{equation}
\label{ordres0}
\ds g(T(h))f(x)= \frac{1}{s_0!}(T(h)-1)^{s_0}f(x)\frac{d^{s_0}}{d \lambda^{s_0}}g(\lambda=1)+..
\end{equation}
et

\begin{equation}
\label{ordreS0}
\ds g(1+h)= \frac{h^{s_0}}{s_0!}\frac{d^{s_0}}{d \lambda^{s_0}}g(\lambda=1)+..
\end{equation}
afin d'arriver \`a

\begin{equation}
\label{onarrive}
{\cal D}_gf(x)= h^{-s_0}(T(h)-1)^{s_0}f(x) = \frac{d^{s_0}}{d x^{s_0}}f(x)
\end{equation}
qui n'existe qu'au cas o\`u $f$ serait $s_0$ fois differentiable de mani\`ere continue. (\ref{onarrive}) est valable si $|\frac{d^s}{d\lambda^s}g(\lambda=1)| < \infty $, $\forall s\in \N_0$. On se rend compte que l'op\'erateur ${\cal D}_g$ de (\ref{onarrive}) est local, il agit seulement sur $f(x)$ au point $x$. Toutefois, la localit\'e de l'op\'erateur ${\cal D}_g$ n'est pas conserv\'ee si la constituante $g$ poss\`ede des d\'eriv\'ees
singuli\`eres au point $\lambda=1$, c'est-\`a-dire s'il existe un ordre $s_1$ tel que
\begin{equation}
\label{ordretelque}
|\frac{d^s}{d\lambda^s}g(\lambda\rightarrow 1)| =\infty \,, s \geq s_1=0,1,2,.. \in N_0
\end{equation}
et $|\frac{d^s}{d\lambda^s}g(\lambda\rightarrow 1)| < \infty$ si $s< s_1$.
Dans ces cas (\ref{limproc2}) devient non-local.
La bri\'evet\'e qui s'impose pour cette communication, nous oblige \`a d\'emontrer seulement que les d\'eriv\'ees fractionnaires sont \'egalement comprises dans la notion de d\'eriv\'ee g\'en\'eralis\'ee (\ref{limproc2}). Supposons la constituante

\begin{equation}
\label{conalph}
g(\lambda)=(\lambda-1)^{\alpha} \,, \hspace{1.5cm} 0<\alpha<1
\end{equation}
avec l'exposant $\alpha$ non-entier. Pour $0<\alpha<1$ la constituante (\ref{conalph})
appartient \`a la cat\'egorie (\ref{exprop1}) et par cons\'equent
(\ref{limproc2}) s'\'ecrit
\begin{equation}
\label{Dfrac}
{\cal D}_gf(x)=h^{-\alpha}\left(T(h)-1\right)^{\alpha}f(x)
\end{equation}
par d\'efinition d'une {\cal d\'eriv\'ee}\footnote{Si $\alpha < 0$ $g$ est du type (\ref{exprop2}) et dans ce cas (\ref{Dfrac}) d\'efinit une {\it int\'egrale} fractionnaire.}  fractionnaire \cite{oldham}.
Nous supposons que la fonction $f(x)$ est choisie telle que $\left(T(h)-1\right)^{\alpha}$
produise une s\'erie convergente soit

\begin{equation}
\label{Dfrac2}
{\cal D}_gf(x)=D_x^{\alpha}f(x)=h^{-\alpha}\sum_{k=0}^{\infty}(-1)^k
\left(\begin{array}{c}\!\!\! \alpha \!\!\! \nonumber \\ \!\!\! k \!\!\! \end{array}\right)f(x-kh)
\end{equation}
o\`u nous avons utilis\'e le fait que $(T(h)-1)^{\alpha}=T(\alpha h)(1-T(-kh))^{\alpha}=(1-T(-kh))^{\alpha}$ car $T(h\alpha)=1$ dans le cas limite $h\rightarrow 0$.
On tient compte de la d\'ecomposition $T(-hk)=(T(-h)-1+1)^k$

\begin{equation}
\label{decompo}
T(-kh)=\sum_{s=0}^{\infty}\left(\begin{array}{c}\!\!\! k \!\!\! \nonumber \\ \!\!\! s \!\!\! \end{array}\right)\left(T(-h)-1\right)^s
\end{equation}

Or il convient d'abord d'\'evaluer la s\'erie d'op\'erateurs

\begin{equation}
\label{Shrel}
S(k_0,h)=\sum_{k=0}^{k_0}(-1)^k\left(\begin{array}{c}\!\!\! \alpha \!\!\! \nonumber \\ \!\!\! k \!\!\! \end{array}\right)T(-kh)=\sum_{s=0}^{\infty}\sum_{k=0}^{k_0}(-1)^k\left(\begin{array}{c}\!\!\! \alpha \!\!\! \nonumber \\ \!\!\! k \!\!\! \end{array}\right)\left(\begin{array}{c}\!\!\! k \!\!\! \nonumber \\ \!\!\! s \!\!\! \end{array}\right)\left(T(-h)-1\right)^s
\end{equation}

qui peut \^etre r\'e-\'ecrite comme

\begin{equation}
\label{interim1}
S(k_0,h)f(x)=\sum_{s=0}^{\infty}A_s(k_0)(T(-h)-1)^sf(x)
\end{equation}

et avec

\begin{equation}
\label{closedformcoeff1}
A_s(k_0)=\sum_{k=0}^{k_0}(-1)^k \left(\begin{array}{c}\!\!\! \alpha \!\!\! \nonumber \\ \!\!\! k \!\!\! \end{array}\right)\left(\begin{array}{c}\!\!\! k \!\!\! \nonumber \\ \!\!\! s \!\!\! \end{array}\right)=
(-1)^{k_0}\left(\begin{array}{c}\!\!\! \alpha-(s+1) \!\!\! \nonumber \\ \!\!\! k_0-s \!\!\! \end{array}\right)
\left(\begin{array}{c}\!\!\! \alpha \!\!\! \nonumber \\ \!\!\! s \!\!\! \end{array}\right) = \frac{-\alpha}{s-\alpha}(-1)^{k_0}\left(\begin{array}{c}\!\!\! \alpha-1 \!\!\! \nonumber \\ \!\!\! k_0 \!\!\! \end{array}\right)\left(\begin{array}{c}\!\!\! k_0 \!\!\! \nonumber \\ \!\!\! s \!\!\! \end{array}\right) \,, \hspace{0.5cm} s=0,1,2,.. \in \N_0
\end{equation}
Alors on trouve pour (\ref{interim1})
\begin{equation}
\label{inter2rim}
S(k_0,h)f(x)= (-1)^{k_0}\left(\begin{array}{c}\!\!\! \alpha-1 \!\!\! \nonumber \\ \!\!\! k_0\!\!\! \end{array}\right)f(x)+\sum_{s=1}^{\infty}A_s(k_0)(T(-h)-1)^sf(x)
\end{equation}
En vue de (\ref{Dfrac2}) nous sommes surtout int\'eress\'es par les repr\'esentations asymptotiques pour $k_0\gg 1$ \`a savoir
\begin{equation}
\label{assalpmun1}
\left(\begin{array}{c}\!\!\! \alpha-1 \!\!\! \nonumber \\ \!\!\! k_0\!\!\! \end{array}\right)(-1)^{k_0} =
\frac{(k_0-\alpha)!}{(-\alpha)!k_0!} \rightarrow \frac{ k^{-\alpha}}{(-\alpha)!} \,\,,\hspace{2cm} k_0\gg1
\end{equation}
et
\begin{equation}
\label{ksurj1}
\left(\begin{array}{c}\!\!\! k_0\!\!\! \nonumber \\ \!\!\! s \!\!\! \end{array}\right) \rightarrow \frac{k_0^s}{s!} \,\,,\hspace{2cm} k_0\gg1
\end{equation}
o\`u $(-\alpha)!$ est d\'efini par (\ref{betafaculty}). Pour $k_0\gg1$ on a donc
\begin{equation}
\label{assympfi1}
A_s(k_0) \rightarrow \frac{-\alpha}{s-\alpha}\frac{k_0^s}{s!}\frac{ k^{-\alpha}}{(-\alpha)!}
\,\,,\hspace{2cm} s=0,1,2,..\infty
\end{equation}
o\`u il faut distinguer les cas $s=0$ et $s>0$ soit\footnote{A remarquer $A_0(k_0\rightarrow \infty)=0$ \'etant une cons\'equence de (\ref{exprop1}) et
et $A_s(k_0\rightarrow \infty)=\infty$ pour $s>0$.}

\begin{equation}
\label{generateres1}
A_s(k_0)=\frac{-\alpha}{(s-\alpha)}
\frac{k_0^{s-\alpha}}{(-\alpha)!s!} =\left\{ \begin{array}{ll} \ds \frac{k_0^{-\alpha}}{(-\alpha)!} & s=0 \nonumber \\ \ds & \nonumber \\
\ds \int_0^{k_0}\frac{k^s}{s!}\frac{(-\alpha)k^{-(\alpha+1)}}{(-\alpha)!}{\rm d}k & s \geq 1
\end{array}\right.
\end{equation}
Il s'ensuit que
\begin{equation}
\label{interimad1}
S(k_0,h)=A_0(k_0)f(x)+\sum_{s=1}^{\infty}A_s(k_0)(T(-h)-1)^s f(x)
\end{equation}
La relation (\ref{interimad1}) est valable aussi pour les fonctions non-diff\'erentiables (h\"old\'eriennes), voire pour les fonctions non-continues.
Ici, il nous faut rendre compte du caract\`ere de la fonction $f$~: si $f(x)$ est infiniment diff\'erentiable (lipschitzienne), on peut poser $T(-h)=e^{-hD_x}$ et dans le cas limite $h \rightarrow 0$
on peut poser $(T(-h)-1)^sf(x)= (-1)^sh^sD_x^sf(x)$. En tenant compte de (\ref{generateres1}) on obtient,
pour $k_0\gg1$, en introduisant la variable continue $0\leq \tau=hk \leq \tau_0=hk_0$~:
\begin{equation}
\label{interimad2}
S(k_0,h)= h^{\alpha}\left\{ \frac{\tau_0^{-\alpha}}{(-\alpha)!} f(x)+ \sum_{s=1}^{\infty}(-1)^sD_x^sf(x)\int_0^{\tau_0}\frac{\tau^s}{s!}\frac{(-\alpha)\tau^{-(\alpha+1)}}{(-\alpha)!}{\rm d}\tau \right\}
\end{equation}
On s'aper{\c c}oit que la somme \`a partir de $s=1$ est une convolution de la s\'erie de Taylor
de $f(x-\tau)-f(x)$. On peut donc r\'e-\'ecrire (\ref{interimad2}) comme
\begin{equation}
\label{interimad3}
S(k_0,h)= h^{\alpha}\left\{ \frac{\tau_0^{-\alpha}}{(-\alpha)!} f(x)+ \int_0^{\tau_0}\left(f(x-\tau)-f(x)\right)\frac{(-\alpha)\tau^{-(\alpha+1)}}{(-\alpha)!}{\rm d}\tau \right\}
\end{equation}
(\ref{interimad3}) est valable pour $k_0\gg1$ avec $\tau_0(h)=hk_0$. Le cas limite $h\rightarrow 0$
de (\ref{interimad3}) qui d\'etermine (\ref{Dfrac2}) d\'epend du choix de $\tau_0(h)$ \`a condition que $k_0 \gg0$.
Il faut souligner que dans (\ref{interimad2}), (\ref{interimad3}) la s\'equence des processus limites joue~:
\begin{equation}
\label{caslimdif}
\lim_{h\rightarrow 0}\left(\lim_{k_0\rightarrow\infty}S(k_0,h)\right)\neq \left(\lim_{h\rightarrow 0}\,S(k_0(h))\right) \,,\hspace{2cm}  {\rm avec} \lim_{h\rightarrow 0} k_0(h)\rightarrow \infty
\end{equation}
Pour cette raison, il y a autant de d\'efinitions diff\'erentes de la d\'eriv\'ee fractionnaire dans la lit\'erature (e.g. \cite{oldham}).

Consid\'erons le cas du membre gauche de (\ref{caslimdif}) \`a savoir $k_0\rightarrow \infty$ alors que $h$ est infinitesimal et fixe. On a donc dans ce cas toujours comme limite sup\'erieure d'int\'egration $\tau_0=hk_0\rightarrow \infty$
dans (\ref{interimad3}) et on obtient
\begin{equation}
\label{Dalph2}
D_x^{\alpha}f(x)= h^{-\alpha}S(k_0=\infty,h)= \int_0^{\infty}\left(f(x-\tau)-f(x)\right)\frac{(-\alpha)\tau^{-(\alpha+1)}}{(-\alpha)!}{\rm d}\tau
\end{equation}
qui peut \^etre r\'e-\'ecrit comme
\begin{equation}
\label{Dalph3}
D_x^{\alpha}f(x)= \frac{(-\alpha)}{(-\alpha)!}\int_{-\infty}^{x}(x-t)^{-(\alpha+1)}\left(f(t)-f(x)\right){\rm d}t
\end{equation}
Les expressions (\ref{Dalph2})-(\ref{Dalph3}) sont connues comme {\it d\'eriv\'ees fractionnaires de Marchaud}\footnote{Malheuresement il existe une certaine confusion sur cette d\'enomination dans la lit\'erature. Nous avons adopt\'e celle de \cite{ross}, mais on devrait y trouver des d\'enominations diff\'erentes.} (Eq. (12) dans \cite{ross}).
On se rend compte que (\ref{Dalph2}) n'existe qu'aux cas o\`u $|f(x-\tau)-f(x)|$ se comporte pour $\tau\rightarrow 0$ comme $C\tau^{\delta}$ avec $\delta > \alpha$, c'est-\`a-dire si $f(x)$ est
h\"old\'erienne, son exposant doit \^etre tel que $0<\alpha < \delta \leq 1$ o\`u le cas $\delta=1$ correspond aux fonctions traditionnelles lipschitziennes.
On a la propriet\'e d\'esir\'ee que $D_x^{\alpha}(const)=0$.
A condition que $f(x)$ soit une fois diff\'erentiable, on arrive \`a
\begin{equation}
\label{Dalph5}
D_x^{\alpha}f(x)=D_x^{\alpha -1}(D_xf(x))= D_x(D_x^{\alpha -1}f(x))=D_x\int_0^{\infty}\frac{\tau^{-\alpha}}{(-\alpha)!}f(x-\tau){\rm d}\tau
\end{equation}
o\`u l'int\'egrale correspond tout \`a fait \`a ce qu'on d\'eduit pour $D_x^{\alpha-1}f(x)$
\`a partir de
(\ref{limproc2})\footnote{Dans notre classification $D_x^{\alpha-1}f(x)$ est une {\it int\'egrale} car $\alpha -1<0$ et correspond \`a l'int\'egrale fractionnaire d'ordre $\alpha-1$ de Riemann-Liouville  \cite{miller}.}.
Afin d'obtenir (\ref{Dalph2}), (\ref{Dalph3}), on a suppos\'e que $f(x)$ remplit les conditions suivantes

\begin{equation}
\label{cond12}
\begin{array}{l}
\ds \lim_{\tau\rightarrow 0}|f(x)-f(x-\tau)| \leq C_0(x) \tau^{\delta}   \nonumber \\  \nonumber \\
\ds \lim_{\tau\rightarrow \infty}|f(-\tau)| \leq C_{\infty} \tau^{\beta}
\end{array}
\end{equation}
avec $\beta < \alpha < \delta \leq 1$. Les relations (\ref{Dalph2}), (\ref{Dalph3}) sont donc plus g\'en\'erales que
(\ref{Dalph5}) qui exige que $f$ soit une fois diff\'erentiable et par cons\'equent $\delta=1$.
Nous rappelons que nous avons toujours suppos\'e que $0<\alpha<1$.
On observe que $f(x)=e^{\lambda x}$ ($\lambda > 0$, pour la convergence\footnote{$Re(\lambda) > 0$ si $\lambda \in \C$.}) est une fonction propre de l'op\'erateur $D_x^{\alpha}$ \`a savoir
\begin{equation}
\label{expfoprop}
\begin{array}{l}
\ds D_x^{\alpha}e^{\lambda x}=h^{-\alpha}(1-T(-h))^{\alpha}e^{\lambda x}=h^{-\alpha}(1-e^{-h\lambda})^{\alpha}e^{\lambda x}=\lambda^{\alpha}e^{\lambda x} \nonumber \\ \nonumber \\
\ds D_x^{\alpha}e^{\lambda x}= \lambda^{\alpha}e^{\lambda x}\int_0^{\infty}\frac{t^{-\alpha}}{(-\alpha)!}e^{-t}{\rm d}t= \lambda^{\alpha}e^{\lambda x}
\end{array}
\end{equation}
Le fait que $e^{\lambda x}$, qui est fonction propre de
l'op\'erateur de translation $T(-h)e^{\lambda x}=e^{\lambda (x-h)}=e^{-h\lambda}e^{\lambda x}$, reste aussi une fonction propre
de $D_x^{\alpha}$ de (\ref{Dalph3}), (\ref{Dalph5}) r\'eside dans le fait que ces deux op\'erateurs
agissent dans le m\^eme espace de fonctions d\'efini sur le domaine $-\infty < x < \infty$\footnote{Ce fait se traduit par la commutativit\'e des deux op\'erateurs $T(h)$ et $D_x^{\alpha}$.}.

\section{Conclusions}
Nous avons introduit le concept d'une d\'eriv\'ee (et int\'egrale) g\'en\'eralis\'ee qui est susceptible de reproduire les d\'eriv\'ees fractionnaires, ce que nous avons d\'emontr\'e \`a l'aide de la d\'eriv\'ee fractionnaire de Marchaud (eqs. (\ref{Dalph2})-(\ref{Dalph5})). De telles d\'eriv\'ees peuvent \^etre d\'efinies par des cas limites \`a partir de la relation (\ref{limproc2}) ou par des relations semblables l\'eg\`erement modifi\'ees. Dans certains cas on d\'eduit des convolutions permettant de capturer des effets non-locaux dans l'espace ou dans le temps.
Par exemple, ce concept permet d'aborder certains syst\`emes m\'ecaniques comportant des interactions inter-particulaires \`a longue distance comme il s'en produit dans les milieux ayant une microstructure sans \'echelle (auto-similaires) \cite{michel-phys_rev_e}.


\begin{thebibliography}{100}
\footnotesize

\bibitem{ross} B. Ross, S.G. Samko, E. Russel Love, Functions that have no First Order Derivative might have Fractional Derivatives of all Orders Less than One, Real Analysis Exchange 20(2) (1994/5), 140-157.

\bibitem{michel-phys_rev_e} T. M. Michelitsch, G. A. Maugin, F. C. G. A. Nicolleau, A. F. Nowakowski, and S. Derogar, Dispersion relations and wave operators in self-similar quasicontinuous linear chains, Phys. Rev. E 80, 011135 (2009).

\bibitem{miller} Miller, K. S. and Ross, B. An Introduction to the Fractional Calculus and Fractional Differential Equations. New York: Wiley, 1993.

\bibitem{oldham} Oldham, K. B. and Spanier, J. The Fractional Calculus: Integrations and Differentiations of Arbitrary Order. New York: Academic Press, 1974.

\bibitem{riesz} Riesz, Marcel (1949), L'int\'egrale de Riemann-Liouville et le probl\`eme de Cauchy, Acta Mathematica 81: 1223, doi:10.1007/BF02395016, MR0030102, ISSN 0001-5962.

\bibitem{samko} S. Samko, A. Kilbas and O. Marichev, Fractional Integrals and Derivatives: Theory and Applications, Gordon and Breach, London (1993).

\end{thebibliography}
\end{document}